\documentclass[11pt,dvipsnames]{article}
\usepackage[style=philosophy-classic,natbib=true,doi=false]{biblatex}
\bibliography{references.bib}
\usepackage{enumerate,amsmath,amssymb,mathrsfs,nicefrac,csquotes,lmodern,setspace,xcolor,ebgaramond-maths}
\usepackage[affil-it]{authblk}
\usepackage[american]{babel}
\usepackage[colorlinks,allcolors=MidnightBlue]{hyperref}

\title{\bf Incomplete yet existent objects: a Nuclear Meinongian approach to quantum metaphysical indeterminacy}
\author[1,2]{Raoni Arroyo}
\affil[1]{Universidade Federal de Santa Catarina (UFSC), Brazil}
\affil[2]{Centre for Logic, Epistemology and the History of Science (CLE), Brazil}
\author[3]{Renato Semaniuc Valvassori}
\affil[3]{Universidade Estadual de Campinas (UNICAMP), Brazil}
\date{Forthcoming in \href{https://periodicos.ufsc.br/index.php/principia}{Principia}\\\today}

\begin{document}
\sloppy
\raggedbottom
\maketitle
\doublespacing

\begin{abstract}
\noindent This article proposes a reading of quantum metaphysical indeterminacy from the perspective of Parsons' Nuclear Meinongianism. In doing so, we identify a fundamental incompatibility between a key feature of Parsons' theory and standard quantum mechanics. Our approach interprets quantum indeterminacy as property incompleteness. However, this move, when combined with Parsons' framework, leads to what we term the ``Incompleteness-Entails-Nonexistence Principle'' (IENP), which implausibly entails the nonexistence of quantum objects. For Meinongianism to be a suitable tool for the metaphysics of quantum mechanics, this principle must be addressed. We argue for dropping the IENP and discuss the resulting metaphysical and metametaphysical consequences.
Nuclear Meinongianism.
Quantum metaphysical indeterminacy.
Standard non-relativistic quantum mechanics.
Toolbox approach to metaphysics
\end{abstract}

\section*{Introduction}

This manuscript proposes a new interpretation of quantum metaphysical indeterminacy using the framework of Parsons' Nuclear Meinongianism. We advance the following argument: (i) quantum indeterminacy entails incompleteness; (ii) in Parsons' view, incompleteness entails nonexistence; yet (iii) quantum indeterminacy does not entail nonexistence. This yields a contradiction, and we argue that a naturalistic methodology, which prioritizes scientific findings, compels us to reject the second premise.

It is widely accepted that standard non-relativistic quantum mechanics exemplifies a case of genuine metaphysical indeterminacy, \textit{viz.}, that there are situations in which there's no fact of the matter about certain properties of quantum objects. Section \S~\ref{sec:quantum} deals with this. 

A metametaphysical trend called the ``Toolbox approach'' to metaphysics states that physics alone doesn't tell us how to metaphysically interpret some of its concepts, so we should learn about them in the literature on metaphysics (\cite{french-mckenzie2012,}; \cite{french2018jgps}). We argue that quantum metaphysical indeterminacy is one such case. To our knowledge, Meinongianism has not been used to interpret quantum metaphysical indeterminacy, so we propose to explore its application here. The approach seems natural, as Parsons' version of Nuclear Meinongianism deals with indeterminacy precisely in terms of incompleteness. However, it is \textit{also} widely accepted that completeness, \textit{viz.}, the fact that every property is determinate, is a feature of all existent objects; hence, only nonexistent objects can be incomplete in this sense. Section \S~\ref{sec:metaontology} deals with that.

With these two common assumptions in both physics and philosophy, one apparently gets the dilemma: either quantum objects are to be understood in existent/nonexistent terms, or quantum objects are a case of existent but incomplete objects. A way out of the dilemma is to stick with physics. The nonexistence of quantum objects contradicts empirical data, so the connection between incompleteness and nonexistence must be dropped. Section \S~\ref{sec:dilemma} deals with this.

It resulted that these two common assumptions (\textit{quantum indeterminacy} in the philosophy of standard quantum mechanics, and the \textit{Incompleteness-Entails-Nonexistence Principle} of Nuclear Meinongian metaontology) cannot go together. Following the meta-Popperian methodology, in case metaphysics and science clash with each other, we should side with science \parencite{arenhartarroyo2021veri}. Hence, the need to modify Parsons' Nuclear Meinongianism if  one wishes to remain Nuclear Meinongian for independent reasons.  

\section{Quantum metaphysical indeterminacy}\label{sec:quantum}

As \textcite{torza2022} nicely puts it, there are several places in which metaphysical indeterminacy might arise:

\begin{quote}
    \textelp{} (i) the `fuzzy' objects of the macroscopic world, such as clouds, mountains and persons; (ii) future contingents and the open future; and (iii) quantum indeterminacy. Putative instances of iii include (iii.a) the failure of value definiteness of quantum observables; (iii.b) the vague identity of quantum objects; and (iii.c) the count indeterminacy arising in quantum field theory. \parencite[p.~338]{torza2022}.
\end{quote}

With the help of this landscape, we may now narrow down and specify our focus even more: according to this taxonomy, we'll focus solely on item ``iii.a'', for which there are two arguments: superposition and contextuality. Let us see them briefly.

\subsection{Superposition}

Einstein, in correspondence with Schrödinger, eloquently presents the issue as follows. Suppose that there's a ball and two boxes. There are not much ways of arranging the situation: either the ball is in box $A$, or in box $B$. A third possibility is the ball being outside the two boxes, but let us bracket this one for now. A quantum-mechanical description of this situation, whenever the boxes are sealed, is by means of superposition, \textit{viz.}, the sum of the ball being in the first box and the ball being in the second box. He then continues:

\begin{quote}
    Now I describe a state of affairs as follows. \textit{The probability is $\nicefrac{1}{2}$ that the ball is in the first box.} Is this a complete description?\\\textit{NO}: A complete statement is: the ball \textit{is} (or is not) in the first box. That is how the characterization of the state of affairs must appear in a complete description.\\\textit{YES}: Before I open them, the ball is by no means in \textit{one} of the two boxes. Being in a definitive box only comes about when I lift the covers.\footnote{The original last sentence (in German) is: ``\textit{Dies Sein in einer bestimmten Shactel kommt erst dadurch zustande, das ich den Deckel aufklappe}.''} 
    \parencite[Einstein, Letter to Erwin Schrödinger, June 19, 1935. Extracted from and translated by][p.~69, original emphasis]{fine1986}.
\end{quote}

The last sentence is crucial. If quantum mechanics completely describes the situation---\textit{i.e.}, if it encodes thoroughly the state of affairs of the physical situation---then the location of the ball is (in this situation) ontologically indeterminate. Einstein found it unacceptable. In fact, these correspondences resulted in the famous article with the Schrödinger's Cat thought experiment \parencite{schrodinger1935orig}. Schrödinger depicted a situation in which even a macroscopic biological system (a cat), if described completely by quantum mechanics, cannot be ascribed a definite value for the property of its life status.

Contrast this with a so-called ``natural view'' according to which ``[every] physical object is determinate in all [determinable] respects, it has a perfectly precise colour, temperature, size, etc.'' \parencite[p.~59]{armstrong1961}. Physical objects described by quantum mechanics---microscopic or macroscopic---should have all determinable properties (\textit{e.g.} location) with determinate and unique values (\textit{e.g.} being located in the first box, or having an `alive' life status). Yet this is precisely what they lack in certain quantum-mechanical situations. This is not to say that such indeterminacy is epistemic---\textit{viz.}, they have a determinate location at all times and we simply fail to know which---or semantic---\textit{viz.}, they have a determinate location but we fail to specify what the object is. In standard quantum mechanics, there is no fact of the matter about the determinate values for a physical system's determinable properties. This is the lack of \textit{value definiteness}: the principle that physical systems have well-definite values for all of their properties at all times. As \textcite[79]{albert1992} emphasizes, the standard way of thinking about quantum mechanics interprets superpositions as ``a state in which there is no matter of fact'' about whether or not objects possess determinate values for such properties.

Many authors have described this quantum indeterminacy as \textit{metaphysical} (or \textit{deep}), \textit{viz.}, indeterminacy \textit{in the world} \parencite[see][Chap.~4 for overview and references]{torza2023}. In the next section, we'll discuss different ways, in the domain of metaphysics, in which such a quantum metaphysical indeterminacy has been cashed out.

\subsection{Contextuality}\label{sec:contextuality}

So far, the discussion of quantum metaphysical indeterminacy has been framed in terms of superpositions. In such cases, indeterminacy arises only in particular quantum-mechanical situations---namely, when a system is prepared in a superposed state and described as lacking a determinate value for some observable until measurement. Yet the phenomenon is in fact deeper. Quantum indeterminacy does not merely occur in the special circumstances of superposition, but is also a consequence of the very structure of the theory itself. This is where \textit{contextuality} kicks in.

According to the Kochen--Specker theorem, it is impossible to assign determinate values to all observables of a quantum system in a way that preserves the functional relations between them \parencite{kochen-specker1967}. Any attempt to uphold global \textit{value definiteness}---the principle that physical systems have well-definite values \textit{for all of their properties at all times}---leads to contradiction within the Hilber-space structure. Importantly, this result is independent of the state of the system or of any particular measurement. Contextuality thus establishes a structural form of indeterminacy: no consistent, non-contextual map from observables to definite values exists for quantum systems.

To see how this works, we'll reconstruct the theorem with the proof of \textcite{peres1991} in non-technical terms. Consider a set of $33$ spin vectors for a spin-$1$ particle. It's possible to arrange these vectors in a way that some of them are mutually orthogonal (i.e., at $90^\circ$ to each other). For example, a group of three vectors might be along the $x$, $y$, and $z$ axes. We can say the spin value along each of these axes is either $0$ or $1$. A key quantum rule is that for any three mutually orthogonal spin vectors, the sum of their squared values must equal $2$. For instance, if we measure the spin in the $x$, $y$, and $z$ directions, the values must be $(0, 1, 1)$, $(1, 0, 1)$, or $(1, 1, 0)$. There are $16$ such groups of three mutually orthogonal vectors in our set of $33$. The theorem shows that we cannot assign a definite $0$ or $1$ value to all $33$ vectors simultaneously while respecting this rule. The argument is that if we sum up the values of all $16$ groups, the total must equal $16 \times 2 = 32$. However, some vectors are part of multiple groups. When we count up how many times each of the $33$ vectors appears in the $16$ groups, we'll find that each one appears an even number of times. This means that the total sum must be an even number, because it's the sum of a bunch of even numbers. This creates a contradiction, as $32$ is an even number, but the sum of the individual values must also be an even number. This proves that we can't consistently assign a definite value to every one of the $33$ properties at the same time. This is often called a `value assignment' proof and shows the impossibility of a non-contextual version of quantum mechanics.

In contrast with indeterminacy due to superpositions, contextuality is not a feature of specific situations, but a general constraint on the ontology of quantum systems. No matter the preparation or the measurement setting, the demand that every observable has a determinate value `at all times'---\textit{viz.}, the demand for global value definiteness---cannot be met. As \textcite[\S~11.2, original emphasis]{lombardiforthcoming} writes, ``\textit{[c]ontextuality} prevents the simultaneous assignment of determinate values to all the properties of a quantum system. Thus, it is in conflict with the principle of \textit{omnimode determination}.''

\section{Indeterminacy, incompleteness, and nonexistence}\label{sec:metaontology}
As we have seen, standard quantum mechanics leads to metaphysical indeterminacy. And this is called quantum metaphysical indeterminacy. Yet, physics textbooks do not specify what such indeterminacy is metaphysically, or how we ought to understand it.

According to the Toolbox Approach to metaphysics, as first developed by \textcite{french-mckenzie2012}, we shouldn't expect to extract such a clear picture from science itself. Instead, we should ``engage with extant metaphysics, draw on the tools it has already developed \textelp{} to help us understand what it is that science is telling us'' \parencite[p.~405]{french2018rout}.

Thereby, metaphysical theories should be used by the philosophy of science in order to interpret science in cases where the latter is silent. And metaphysicians have indeed worked their way through quantum metaphysical indeterminacy. A comprehensive survey of extant positions may be found in \textcite[pp.~11193--11209]{fletcher-taylor2021}. In the present paper, we aim to assess what may at first appear to be a promising addition to the metaphysician's toolbox: a Meinongian approach to quantum metaphysical indeterminacy.  In order to do so, let's recall how (metaphysical) indeterminacy is defined:

\begin{quote}
    Indeterminacy is the situation in which an object has a determinable property, but no determinate value for that determinable. \parencite[p.~76]{plewis2016}.
\end{quote}

We must keep in mind that property-completeness is traditionally thought to be a feature of all objects---a thesis supported by the strong intuition we have according to which reality can have no gaps. It is even hard for us to think of incomplete objects: how could something be neither blue nor non-blue; neither round nor non-round, or neither alive nor non-alive? If an object is not blue (let us say it is red), we automatically allow the inference that leads us to assume that it is \textit{non-blue}.

In at least some conceivable scenarios, however, quantum metaphysical indeterminacy can be thought of in terms of \textit{property-incompleteness}. Recall Einstein's `ball-in-a-box', or `Schrödinger's cat' thought experiments. But that's abstruse. Because, when it comes to the property of \textit{having a certain life status} \parencite{calosi-wilson2018}, our intuitions are even stronger: if something is not alive, we would not think twice to infer it is non-alive (or even dead if it is an animal, like a cat). The traditional philosopher would say that it is impossible for an object to be incomplete, and in most cases, assuming completeness as a feature of everything---recall, that's what \textcite{armstrong1961} had in mind---will not harm any metaphysical enterprises. 

The only philosophical perspective we are aware of that challenges such a widespread assumption concerning property-completeness is Meinongianism. Developed by Alexius Meinong and embraced by contemporary philosophers such as Graham Priest, Edward Zalta, and Terence Parsons, Meinongianism, as a metaontology, is characterized by the recognition of the possibility of quantifying over objects that do not exist, in such a way that these nonexistent objects (despite their nonexistence) can have ontological and/or semantic roles (see \cite[chapter 9]{Correia2012-CORMGU}). As we will see, it is precisely the acceptance of nonexistents in the general domain of quantification that allows certain Meinongian trends to reject the thesis according to which all objects are complete. And, for our purposes here, we'd like to highlight the connection between metaphysical incompleteness and nonexistence according to Meinongianism: the former as a necessary condition for the latter.

A common example of how nonexistent objects could be incomplete is by considering the case of fictional characters. Take, for instance, Gandalf from J. R. R. Tolkien's works. Given that Gandalf is a fictional object, it is intuitive to say that he does not exist (after all, he is not part of the space-time and seems to have no real causal powers). Meinongians, in general, keep this intuition while at the same time accepting that we can truly predicate a series of properties of Gandalf. Considering what is explicitly detailed in Tolkien's stories, we can confidently assign him, for example, the properties of \textit{being-a-wizard} and \textit{inhabiting-the-Middle-earth}. On the other hand, since Tolkien's writings never specify the exact number of beard hairs on Gandalf's face, Meinongians allow for the possibility of Gandalf possessing neither the property of ``having exactly 100,000 beard hairs,'' nor the property of ``not having exactly 100,000 beard hairs.'' Here we can see an explicit case of property-incompleteness. 

Now let us see how this notion of property-incompleteness fits into Parsons' Nuclear Meinongian theory. According to his version, for every set of nuclear properties, there is a corresponding object (whether existent or nonexistent). In order to avoid the classic Russellian objections to Meinongian theories of objects (namely, that it violates the principles of non-contradiction and excluded middle, and, most importantly, that it is trivialized by Russel's Comprehension Principle, see \cite[chapter 7]{Berto2015-BEROAM-2}), Parsons makes this distinction between nuclear properties---understood as ordinary characterizing properties, such as \textit{being-round}, \textit{being-alive} or \textit{being-red}---and extra-nuclear properties---those properties that determine the objects' ontological and technical status (more precisely, extra-nuclear properties can determine not only the ontological and technical status of an object but also its modal and intentional status, see \cite[p.~39]{parsons1980}), such as \textit{being-existent, being-complete or being-ficitional}. With this terminology in hand, Parsons says:
\begin{quote}
    By calling an object `complete,' I mean that for any \textit{nuclear} property, the object has either that property or it has its negation. \parencite[p.~19, emphasis added]{parsons1980}.
\end{quote}

An incomplete object is then, by definition, an object that does not possess certain properties or their negation. To be sure, let us check the following example to see how Parsons uses the concept of being complete:

\begin{quote}
    Consider the object whose properties are goldness and mountainhood. It does not have the property of blueness, nor does it have the property of nonblueness; I will say that it is indeterminate with respect to blueness. That object will in fact be indeterminate with respect to every nuclear property except goldness and mountainhood. \textelp{A}dd to the ``the gold mountain'' all nuclear properties that are entailed by goldness and mountainhood. Then it will have, for example, the nuclear property of either-being-located-in-North-America-or-not-being-located-in-North-America, but it will not have either of those disjuncts; it will be indeterminate with respect to being located in North America. \parencite[pp.~20--21]{parsons1980}.
\end{quote}

For our purposes, however, the most important thing to keep in mind is that, according to \textcite[p.~20]{parsons1980}, ``all existing objects are complete.'' He does not even argue for this thesis: he just accepts it, respecting our metaphysical intuition of what seems to be obviously true. In other words, even if some objects are incomplete, as Parsons claims to be, all incomplete objects would be nonexistent and, therefore, not part of spatiotemporal reality---albeit part of the \textit{extended} notion of ``reality'' as per the Meinongian view, which encompasses nonexistent objects. As another example of that attitude, \textcite{reicher2022} categorically states, in her SEP entry on \textit{Nonexistent Objects}, that ``[i]ncomplete objects are necessarily nonexistent.'' Let us call this the ``Incompleteness-Entails-Nonexistence Principle'' (IENP). The IENP acts as a transitivity between indeterminacy and inexistence. Something along the following lines.

\begin{itemize}
    \item[1.] Indeterminacy $\implies$ Incompleteness;
    \item[2.] Incompleteness $\implies$ Nonexistence;
    \item[$\therefore$] Indeterminacy $\implies$ Nonexistence;
\end{itemize}

The IENP entails that an incomplete object should count as nonexistent. This might be fair to some kinds of nonexistent objects, such as fictional ones. After all, Gandalf is incomplete with regards to some properties, \textit{e.g.} the color of his socks, and---most of all---lacks both exact \textit{and} weak location. This is all fine, as he's a nonexistent (\textit{qua} fictional) object. 

Parsons establishes that for every set of nuclear properties, there is an object corresponding to it. The properties whose set corresponds to a certain object are all the properties the object has (call this principle the \textit{Object Principle}). That is in accordance with the intuition we have that we can arbitrarily stipulate a set of properties to designate an object and then have this object as a target of our intentional states---precisely what authors of fictional stories do all the time (they stipulate a name followed by a series of arbitrary predicates and start to designate the object satisfying these predicates as a determinate character of one such stories).

According to Parsons, all objects having the extra-nuclear property of existence are complete and logically closed. Nevertheless, it is still possible for nonexistent objects to be complete: think of a nonexistent object that has all the properties the actual Robert De Niro has, except for having a golden front tooth, instead of a natural one. That is an example of a nonexistent complete object---and it is easy to come up with an infinite list of such complete-yet-nonexistent objects \textit{e.g.}, the right-handed Kurt Cobain, \textit{etc.} In other words, according to Parsons' approach, completeness cannot be used as a criterion to identify existent objects.

To our best knowledge, neither \textcite{parsons1980}, \textcite{reicher2022}, nor anyone else has provided a clear and explicit argument \textit{for} the IENP. Nevertheless, we believe that the point of the IENP is an appeal to intuition, as if it were a kind of `basic metaphysical law'---something almost self-evident, such that no one would need a philosophical argument to be convinced of its truth. And perhaps this is why we find full-fledged arguments for it scarce in the literature. Intuition seems to give us \textit{prima facie} motivation to introduce the IENP---and few people would argue otherwise, if it weren't for the empirical findings of quantum mechanics, \textit{viz.} quantum metaphysical indeterminacy. Perhaps something similar happened with Leibniz's Law (the principle of the Identity of Indiscernibles). Until counterexamples from quantum mechanics emerged \parencite[\textit{viz.} the argument from Bose--Einstein statistics, see][]{frenchkrause2006}, it seemed reasonable to appeal to intuition when defending the principle.

But for \textit{concreta} like electrons, measuring devices, observers' brains, or cats, `incompleteness-plus-nonexistence' is highly implausible. The next section tackles this problem.

\section{Incomplete yet existent}\label{sec:dilemma}
Equipped with these metaphysical tools, let us return to standard non-relativistic quantum mechanics. As we saw, in the standard way of thinking about quantum-mechanical descriptions, there are familiar situations in which quantum objects are incomplete in this sense. According to Parsons' Nuclear Meinongianism \parencite[\textit{e.g.},][]{parsons1980}, this means that quantum objects in this situation are nonexistent objects. This is surely odd, but it's where we ended up by considering the implications of both standard quantum mechanics and Parsons' Nuclear Meinongianism at face value. So let's take a closer look.

It took us three premises to arrive at where we are now:
\begin{enumerate}
    \item[P.~1] Standard quantum-mechanical objects are metaphysically indeterminate with regards to its full set of properties.
    \item[P.~2] Metaphysically indeterminate objects are incomplete objects; metaphysically determinate objects are complete objects.
    \item[P.~3] IENP: \textit{All} incomplete objects are necessarily nonexistent objects; completeness is a necessary (albeit not sufficient) condition to existence.
\end{enumerate}

Of course, one might challenge premise P.~1 on the grounds that there's no metaphysical indeterminacy in different interpretations/reconstructions of quantum mechanics \parencite{glick2017}. However, if we focus exclusively on \textit{standard} quantum mechanics \parencite[\textit{e.g.,}][]{textbookQM1}, then P.~1 is the case, as the Kochen-Specker theorem is proved within the standard Hilbert-space formalism. For joint determination of multiple determinables, \textit{e.g.} position and momentum, or spin values in all directions, as this is impossible due to no-go theorems such as Kochen--Specker's; for a single determinable, such as location, P.~1 holds in certain situations due to the EEL.

P.~2 also seems to be the case, as we argue it in three steps. First, recall Lewis'  definition of \textit{metaphysical indeterminacy}: ``\textelp{} an object has a determinable property, but no determinate value for that determinable'' \parencite[p.~76]{plewis2016}. Then, compare that with Parsons' definition of a ``complete'' object: ``\textelp{it} has either that property or it has its negation'' \parencite[p.~19]{parsons1980}. Now, in standard-quantum-mechanical situations of superposition such as those captured by the scope of the thought experiments by Einstein and Schrödinger, we find it hard (if not impossible!) to assign a determinate value to the position determinable; or to say that the quantum object has the property of being located in the first box or having a life status, say, $A$; nor of being located/having a life status $\neg A$ due to indeterminacy stemming from superposition. Due to contextuality, the determinables position and momentum cannot be both simultaneously determinate (because since the Kochen--Specer theorem, the Heisenberg Principle cannot be interpreted in merely epistemic terms, but must be understood as positing an ontological claim). Hence, quantum metaphysical indeterminacy and Parsons' Nuclear Meinongian completeness fit the situation \parencite[see also][]{arroyo2022}.

That said, let us move on to premise P.~3, the IENP, which is the one that generates trouble.
\begin{description}
    \item[\bf Accept IENP.] Due to the implications of the standard quantum-mechanical formalism, quantum objects don't exist.
    \item[\bf Deny IENP.] Standard quantum-mechanical objects aren't nonexistent objects; they're existent, but incomplete objects.
\end{description}

According to \textcite[p.~358]{jacquette2015}, one dire consequence of the assumption that quantum objects are incomplete, along with the IENP, is ontological nihilism. This implication arises from an argument based on the principle of supervenience. If macroscopic objects supervene on quantum objects, it becomes seemingly implausible to assert that existence supervenes on nonexistence. Consequently, this reasoning suggests that macroscopic objects---measuring apparatuses, cats, human beings---may not exist at all. It's a curious outcome, yet it emerges as a logical entailment when we scrutinize the implications of quantum metaphysical indeterminacy under the IENP.\footnote{\textcite{jacquette2015} offers a different perspective by denying that quantum objects are incomplete, presenting reasons that are beyond IENP, but the conditional implication is what concerns us.} Under the IENP, ontological nihilism is a natural result once the completeness of quantum objects is questioned, that is, if they exemplify metaphysical indeterminacy.

This seems to put us in the face of a dilemma, but there's none. If we want to take physics seriously, this is not a dilemma because we just can't accept the IENP. It's against the experimental data. Consider once a typical double-slit setup: when both slits are open, we cannot help but describing the situation as a superposition of the quantum object being located at slit $A$ plus being located at region $\neg A$. This doesn't---cannot!---mean that the quantum object \textit{is} located at $A$, nor at $\neg A$ due to a (textbook) a phenomenon called \textit{interference} \parencite{albert1992}; similarly, this doesn't mean that the object is multilocated at both $A$ and $\neg A$ due to well-known violations of the Born rule \parencite{calosi2022}; finally, this doesn't mean also that the object is in neither regions, as we know for an experimental fact that when slits are both closed there is no detection \parencite{plewis2016}. \textcite{albert1992} summarizes:

\begin{quote}
Electrons passing through this apparatus, in so far as we are able to fathom the matter, do not take route $h$ [$A$] and do not take route $s$ [$\neg A$] and do not take both of those routes and do not take neither of those routes; and the trouble is that those four possibilities are simply all of the logical possibilities that we have any notion whatever of how to entertain! \parencite[11]{albert1992}
\end{quote}

Be as it may, this also means that the electrons in this situation cannot simply cease to exist, as they have a causal---and empirical!---consequence in the statistical account of hits the detection screen. Hence, IENP cannot be the case due to experimental reasons. Interference grants existence, so to speak. Moreover, the sentence implied by accepting the IENP is, strictly speaking, false when it comes to the impossibility of joint determination of incompatible observables/determinables (\textit{e.g.} position and momentum, spin values in all axes). Accepting the IENP would imply that quantum objects are nonexistent---because incomplete with regards to the joint determination of multiple determinables---even in measurement-like situations, which is highly implausible.

Methodologically speaking, this seems to be an application of ``experimental metaphysics'' \parencite{shimony1984}, such as the ``meta-Popperian'' method \parencite{arenhartarroyo2021meta},\footnote{See also \textcite[pp.~15--16]{arenhartarroyo2021veri} for a brief presentation of the meta-Popperian method, and references therein.} according to which science may help us in narrowing down the possibilities of metaphysical interpretations to scientific concepts. In our case, Parsons' Nuclear Meinongianism---at least as it is, \textit{viz.}, with the IENP---is not compatible with (standard) quantum metaphysical indeterminacy. Hence, Parsons' Nuclear Meinongianism must be either abandoned or at least modified, if one wishes to remain Meinongian for independent reasons (arguably, one might wish to remain metaontologically Meinongian to have a better account of mental intentionality and fictionality, see \cite{Berto2012-BEREAA-2}) while maintaining the standard interpretation of quantum mechanics---but the prospects of having a Meinongian analysis of quantum metaphysical indeterminacy seem doomed. 
as a metaphysical possibility for interpreting quantum metaphysical indeterminacy \textit{or} it should be at least modified.   

Now, how exactly is the IENP crucial for Parsons' Nuclear Meinongianism? And what would it look like if we did away with it? Actually, IENP turns out to be unnecessary for the Parsons' Nuclear Meinongian doctrine. Recall: to be a Nuclear Meinongian is (i) to accept the possibility of quantifying over nonexistent objects, (ii) to recognize that nonexistent objects can have properties just like the existent ones, and (iii) to acknowledge the metaphysical distinction between Nuclear and Extra-Nuclear properties---something crucial to avoid Russellian objections arising from the assumptions (i) and (ii). 

Parsons admits IENP within his theory of objects just in order to accommodate our intuitions concerning metaphysical completeness. The point is: quantum mechanics \textit{is} counterintuitive! That is: our metaphysical intuitions are not always good guides when we are dealing with the fundamentals of physics.

If we look at Sylvan's version of Nuclear Meinongianism, for instance, we see no principle equivalent to Parsons' IENP. In his work \textit{Exploring Meinong's Jungle and Beyond}, \textcite[p.~721]{routley1980} says that (metaphysical) determinacy is not a good criterion for existence, given that ``many objects that exist are not fully determinate'' . He mentions the metaphysical indeterminacy of quantum objects (micro-particles) and even points to the supposed indeterminacy of borders-vague objects such as clouds and waves. 

In other words, Nuclear Meinongianism itself is not incompatible with standard quantum mechanics, but Parson's IENP must be rejected if we want to respect the empirical data and/or the standard way of interpreting quantum mechanics. Nevertheless, there's at least one aspect of Parson's theory that may resist this strike: his analysis of incompleteness. Given that a non-Meinongian wouldn't use his terminology of nuclear/extra-nuclear properties, a seemingly sound and theoretically useful principle inspired on Parsons' account could be formulated as follows:

\begin{description}
    \item[Completeness:] An object is said to be complete if, and only if, for any property, that object has either that property or its negation.
 \end{description}

A supplementary principle that could also be added to help interpret quantum mechanics under the standard framework would be:

\begin{description}
    \item[Determinacy:] An object is metaphysically determinate if, and only if, it is complete. 
\end{description}

\section*{Conclusion}
Using Parsons' Nuclear Meinongianism as a template, we examined whether quantum metaphysical indeterminacy can be modeled as property incompleteness. We argued that the Incompleteness-Entails-Nonexistence Principle (IENP) collides with well-established features of quantum theory. Hence, anyone who seeks to cash out quantum indeterminacy as incompleteness must jettison IENP. For the Meinongian approach to be viable as an interpretive tool, the principle at stake must be rejected. More generally, this illustrates how scientific constraints can and should refine metaphysical theorizing.

\subsection*{Acknowledgments}
Earlier versions of this work were presented at the SNEL seminars (2024, UFSC, Brazil) and at the Workshop de Metafísica de la Ciencia Contemporánea: Nuevas Perspectivas (2025, SADAF, Argentina). This final version has benefited from the inputs of Cristián Soto, Christian de Ronde, Jonas Becker Arenhart, Leandro Beviláqua, Ivan Ferreira da Cunha, and Olimpia Lombardi.
Raoni Arroyo was supported by grants \#$315067$/$2025$-$0$ and \#$446478$/$2024$-$5$, National Council for Scientific and Technological Development (CNPq), call CNPq/MCTI/FNDCT \#$21$/$2024$. The earlier versions of this manuscript were produced during his research visit to the Centre for Logic, Epistemology and the History of Science (CLE), supported by grant \#$2021$/$11381$-$1$, São Paulo Research Foundation (FAPESP), and the Department of Philosophy, Communication and Performing Arts of the Unviersity of Roma Tre, Rome, Italy, supported by grant \#$2022$/$15992$-$8$, São Paulo Research Foundation (FAPESP).
Renato Semaniuc Valvassori was supported by grant \#$2024$/$17481$-$6$, São Paulo Research Foundation (FAPESP). The earlier versions of this manuscript were produced during his research visit to the University of Turin, Turin, Italy, supported by grant \#$2022$/$13928$-$0$, São Paulo Research Foundation (FAPESP).

\printbibliography

@book{Berto2015-BEROAM-2,
	address = {New York},
	author = {Francesco Berto and Matteo Plebani},
	publisher = {Bloomsbury Academic},
	title = {Ontology and Metaontology: A Contemporary Guide},
	year = {2015}
}

@article{arenhartarroyo2021meta,
author={Jonas R. B. Arenhart and Arroyo, Raoni},
title={On physics, metaphysics, and metametaphysics},
journal={Metaphilosophy},
volume={52},
number={2},
doi={10.1111/meta.12486},
pages={175--199},
year={2021}
}

@article{arenhartarroyo2021veri,
author={Jonas R. B. Arenhart and Arroyo, Raoni},
title={{The Spectrum of Metametaphysics: Mapping the state of art in scientific metaphysics}},
journal={Veritas},
volume={66},
doi={10.15448/1984-6746.2021.1.41217},
number={1},
year={2021}
}

@book{armstrong1961,
    author ={Armstrong, David M.},
    title ={{Perception and the physical world}},
    address={London},
    publisher ={Routledge},
    year ={1961}
}

@book{Berto2012-BEREAA-2,
	address = {Dordrecht},
	author = {Francesco Berto},
	editor = {},
	publisher = {Synthese Library, Springer},
	title = {Existence as a Real Property: The Ontology of Meinongianism},
	year = {2012}
}

@article{calosi2022,
  author  = {Calosi, Claudio},
  title   = {There Are No Saints: Or, Quantum Multilocation},
  journal = {Grazer Philosophische Studien},
  volume  = {99},
  pages   = {30--49},
  year    = {2022}
}

@article{calosi-wilson2018,
title={Quantum metaphysical indeterminacy},
author={Calosi, Claudio and Wilson, Jessica},
journal={Philosophical Studies},
volume={176},
doi={10.1007/s11098-018-1143-2},
pages={2599--2627},
year={2019}
}

@book{Correia2012-CORMGU,
	address = {Cambridge},
	editor = {Fabrice Correia and Benjamin Schnieder},
	publisher = {Cambridge University Press},
	title = {Metaphysical Grounding: Understanding the Structure of Reality},
    doi={10.1017/CBO9781139149136},
	year = {2012}
}

@book{fine1986,
    Author={Arthur Fine},title ={The shaky game: Einstein, realism, and the quantum theory},
	Address = {Chicago},
	Publisher = {University of Chicago Press},
	Year = {1986}
}

@article{fletcher-taylor2021,
    author ={Fletcher, Samuel C. and Taylor, David E.},
    title ={Quantum indeterminacy and the eigenstate-eigenvalue link},
    journal ={Synthese},
    volume={199},
    doi={10.1007/s11229-021-03285-3},
    year ={2021}
}

@book{frenchkrause2006,
	Address = {Oxford},
	Author = {French, Steven and Krause, Décio},
	Publisher = {Oxford University Press},
	Title = {Identity in physics: A historical, philosophical, and formal analysis},
	Year = {2006}}

@incollection{french2018rout,
	Address = {New York},
	Author = {Steven French},
	Booktitle = {The Routledge Handbook of Scientific Realism},
	Editor = {Juha Saatsi},
	Pages = {394--406},
	Publisher = {Routledge},
	Title = {Realism and Metaphysics},
	Year = {2018}}

@article{french2018jgps,
	author = {Steven French},
	title = {{Toying with the Toolbox: how Metaphysics Can Still Make a Contribution}},
	journal = {Journal for General Philosophy of Science},
	volume = {49},
	year = {2018},
    doi={10.1007/s10838-018-9401-8},
	pages = {211--230}
}

@article{french-mckenzie2012,
	Author = {French, Steven and M{c}Kenzie, Kerry},
	Journal = {European Journal of Analytic Philosophy},
	Number = {1},
	Pages = {42--59},
	Title = {{Thinking Outside the Toolbox: Towards a More Productive Engagement Between Metaphysics and Philosophy of Physics}},
	Volume = {8},
	Year = {2012}
}

@article{glick2017,
title={Against quantum indeterminacy},
author={David Glick},
journal={Thought: A Journal of Philosophy},
volume={6},
doi={10.1002/tht3.250},
number={3},
pages={204--213},
year={2017}
}

@book{jacquette2015,
    author ={Jacquette, Dale},
    title ={{Alexius Meinong, The Shepherd of Non-Being}},
    publisher ={Springer},
    address={Cham},
    doi={10.1007/978-3-319-18075-5},
    year ={2015}
}

@article{kochen-specker1967,
author={Simon Kochen and E. P. Specker},
title={The Problem of Hidden Variables in Quantum Mechanics},
journal={Journal of Mathematics and Mechanics},
volume={17},
number={1},
pages={59--87},
year={1967}
}

@book{plewis2016,
	Address = {New York},
	Author = {Lewis, Peter J.},
	Publisher = {Oxford University Press},
	Title = {Quantum Ontology},
	Year = {2016}}

@book{lombardiforthcoming,
    author ={Olimpia Lombardi},
    title ={The Modal-Hamiltonian Interpretation of Quantum Mechanics: Making sense of the quantum world},
    publisher ={Oxford University Press},
    year ={forthcoming}
}

@book{textbookQM1,
author={Claude Cohen{-}Tannoudji and Bernard Diu and Franck Laloë},
title={{Quantum Mechanics, Volume I: Basic Concepts, Tools, and Applications}},
translator={Susan Reid Hemley and Nicole Ostrowsky and Dan Ostrowsky},
edition={2nd},
publisher={Wiley},
address={Weinheim},
year={2020}
}

@book{parsons1980,
author={Terence Parsons},
title={{Nonexistent objects}},
publisher={Yale University Press},
address={New Heaven},
year={1980}
}

@InCollection{reicher2022,
	author       =	{Reicher, Maria},
	title        =	{{Nonexistent Objects}},
	booktitle    =	{The {Stanford} Encyclopedia of Philosophy},
	editor       =	{Edward N. Zalta and Uri Nodelman},
	howpublished =	{\url{https://plato.stanford.edu/archives/win2022/entries/nonexistent-objects/}},
	year         =	{2022},
	publisher    =	{Metaphysics Research Lab, Stanford University}
}

@book{routley1980,
	address = {Canberra},
	author = {Richard Routley},
	publisher = {Research School of Social Sciences, Australian National University},
	title = {{Exploring Meinong's Jungle and Beyond: An Investigation of Noneism and the Theory of Items}},
	year = {1980}
}

@article{schrodinger1935orig,
title={{Die gegenwärtige Situation in der Quantenmechanik}},
author={Schrödinger, Erwin},
journal={Naturwissenschaften},
pages={807--812},
doi={10.1007/BF01491891},
volume={23},
year={1935}
}

@article{shimony1984,
    author ={Shimony, Abner},
    title ={{Contextual hidden variables theories and Bell's inequalities}},
    journal ={British Journal for the Philosophy of Science},
    volume={35},
    doi={10.1093/bjps/35.1.25},
    issue={1},
    pages={25--45},
    year ={1984}
}

@incollection{torza2022,
	author = {Alessandro Torza},
	title = {{Derivative Metaphysical Indeterminacy and Quantum Physics}},
	pages = {337--350},
	booktitle = {{Quantum Mechanics and Fundamentality: Naturalizing Quantum Theory between Scientific Realism and Ontological Indeterminacy}},
	editor = {Valia Allori},
	publisher = {Springer},
    address={Cham},
	year = {2022}
}

@book{torza2023,
    author ={Alessandro Torza},
    title ={Indeterminacy in the World},
    publisher ={Cambridge University Press},
    series={Elements in Metaphysics},
    doi={10.1017/9781009057370},
    year ={2023}
}

@book{albert1992,
	Address = {Cambridge},
	Author = {Albert, David Z.},
	Publisher = {Harvard University Press},
	Title = {Quantum mechanics and experience},
	Year = {1992}}

@article{peres1991,
  author       = {Asher Peres},
  title        = {Two simple proofs of the {Kochen--Specker} theorem},
  journaltitle = {Journal of Physics A: Mathematical and General},
  volume       = {24},
  number       = {4},
  pages        = {L175--L178},
  year         = {1991},
  doi          = {10.1088/0305-4470/24/4/003}
}

@article{arroyo2022,
    author ={Raoni Arroyo},
    title ={The Kochen–Specker theorem and ontological (in)completeness of quantum objects},
    journal ={CLE e-Prints},
    year ={2022}
}
\end{document}